\documentstyle[11pt]{article}
\begin{document}

\author{John D. Barrow and Kerstin E. Kunze \\
Astronomy Centre, University of Sussex\\
Brighton BN1 9QH, U.K.}
\title{Exact Inhomogeneous Cosmological Models}
\maketitle
\date{}

\begin{abstract}
We present new exact inhomogeneous vacuum cosmological solutions of
Einstein's equations. They provide new information about the nature of
general cosmological solutions to Einstein's equations.

\end{abstract}

Inhomogeneous and anisotropic solutions of Einstein's equations are of great
interest because of the subtleties of their non-linearity and the clues they
offer about the generic behaviour of cosmological models near spacetime
singularities and at late times. They provide testing grounds for
high-energy physics in the early universe and guide the development of
quantum cosmology. Here, we present a new class of inhomogeneous vacuum
solutions of Einstein's equations which generalise the Einstein-Rosen
solutions [1]. They represent inhomogeneous counterparts of the most general
Bianchi-type vacuum universes and may form a leading-order approximation to
part of a general solution of the vacuum Einstein equations. Moreover,
anisotropic universes containing perfect fluids with pressure less than the
energy density behave in general like vacuum universes at early times.

Consider the generalized Einstein-Rosen metrics: 
\begin{equation}
ds^2=e^{\psi (\xi ,z)}\ (-d\xi ^2+dz^2)+g_{ab}(\xi ,z)dx^adx^b,
\end{equation}
where $\xi $ is a timelike coordinate; $x^a$, $a=1,2$, and $z$ are spacelike
coordinates. These metrics admit an abelian group of isometries, $G_2$.
Spatially-homogeneous cosmologies of Bianchi types I-VII and the
axisymmetric cases of Bianchi types VIII and IX, admit such a $G_2$ and are
particular cases [2]. Properties of (1) depend on whether $B_\mu \equiv
\partial _\mu (det\,g_{ab})^{\frac 12}$ is spacelike, timelike or null
(Greek indices run $0\rightarrow 3$). The cases with a globally null or
spacelike $B_\mu $ correspond to plane or cylindrical gravitational waves,
respectively [2]. Metrics where the sign of $B_\mu B^\mu $ varies throughout
the spacetime describe colliding gravitational waves [3] or cosmologies with
timelike and spacelike singularities [4]. Metrics with a globally timelike $%
B_\mu $ describe cosmological models with spacelike singularities. If the
spacelike hypersurfaces are compact then the allowed spatial topologies [5]
are a 3-torus, $S^1\otimes S^1\otimes S^1,$ for $(detg_{ab})^{\frac 12}=\xi $%
; a hypertorus, $S^1\otimes S^2$, or a 3-sphere, $S^3$, for $%
(detg_{ab})^{\frac 12}=\sin z\sin \xi $ with $0\leq z,\;\xi \leq \pi $. We
shall present solutions for the globally timelike case. Metrics of this form
have been studied approximately by Belinskii and Khalatnikov [6] who argued
that they can provide the leading approximation to a general solution of the
vacuum Einstein equations near the initial singularity.

The metric $g_{ab}$ in (1) can be written as [6] 
\begin{equation}
g_{ab}=R\left( 
\begin{array}{lr}
e^\alpha \cosh \beta & \sinh \beta \\ 
\sinh \beta & e^{-\alpha }\cosh \beta
\end{array}
\right)
\end{equation}
where $\alpha =\alpha (\xi ,z)$, $\beta =\beta (\xi ,z)$, and $R=R(\xi ,z)$,
so $B_\mu \equiv R_{,\mu }$. The vacuum Einstein equations are

\begin{equation}
\ddot R-R^{\prime \prime }=0,
\end{equation}

\begin{equation}
\ddot \alpha +\frac{\dot R}R\dot \alpha -\alpha ^{\prime \prime }-\frac{%
R^{\prime }}R\alpha ^{\prime }=2(\alpha ^{\prime }\beta ^{\prime }-\dot
\alpha \dot \beta )\tanh \beta ,
\end{equation}

\begin{equation}
\ddot \beta +\frac{\dot R}R\dot \beta -\beta ^{\prime \prime }-\frac{%
R^{\prime }}R\beta ^{\prime }=\frac 12(\dot \alpha ^2-\alpha ^{\prime
}{}^2)\sinh 2\beta ,
\end{equation}

\begin{equation}
\ddot \psi -\psi ^{\prime \prime }+\frac 12(\dot \beta ^2-\beta ^{\prime
}{}^2)+\frac 12(\dot \alpha ^2-\alpha ^{\prime }{}^2)\cosh ^2\beta +\frac
12\left( \left( \frac{R^{\prime }}R\right) ^2-\left( \frac{\dot R}R\right)
^2\right) =0,
\end{equation}

\begin{equation}
\frac{\dot R}R\dot \psi +\frac{R^{\prime }}R\psi ^{\prime }-\frac 12(\dot
\beta ^2+\beta ^{\prime }{}^2)-\frac 12(\dot \alpha ^2+\alpha ^{\prime
}{}^2)\cosh ^2\beta +\frac 12\left( \left( \frac{R^{\prime }}R\right)
^2+\left( \frac{\dot R}R\right) ^2\right) -\frac{R^{\prime \prime }}R-\frac{%
\ddot R}R=0,
\end{equation}
where dot denotes $\frac \partial {\partial \xi }$ and a prime denotes $%
\frac \partial {\partial z}$. Physically, these equations describe the
propagation of two inhomogeneous gravitational waves ($\alpha $ and $\beta )$
on an inhomogeneously anisotropic background spacetime. Assume that

\begin{eqnarray}
\beta =f(\alpha ),
\end{eqnarray}
where $f$ is to be determined. Eqns. (4) and (5) yield

\begin{eqnarray}
\left[ \frac{d^2f}{d\alpha ^2}-2\left( \frac{df}{d\alpha }\right) ^2\tanh
f-\frac 12\sinh 2f\right] \left[ \alpha ^{\prime }{}^2-\dot \alpha ^2\right]
=0.
\end{eqnarray}

If $\alpha ^{\prime }{}^2-\dot \alpha ^2\neq 0,$ the general solution is

\begin{eqnarray}
\beta =f(\alpha )={\rm artanh}(Ae^\alpha +Be^{-\alpha })
\end{eqnarray}
with $A$ and $B$ constants. Using (9) in (4)-(5), and defining a new
function $F(\omega )\ $with $\omega \equiv \omega (\xi ,z)$ by $F(\omega
)\equiv e^\alpha ,$ we find two differential equations equivalent to (4) \&
(5): 
\begin{eqnarray}
\ddot \omega +\frac{\dot R}R\dot \omega -\omega ^{\prime \prime }-\frac{%
R^{\prime }}R\omega ^{\prime } &=&0, \\
\frac{d^2F}{d\omega ^2}+\left( \frac{dF}{d\omega }\right) ^2\frac{%
3A^2F^2-B^2F^{-2}-(1-2AB)}{(1-2AB)F-A^2F^3-B^2F^{-1}} &=&0.
\end{eqnarray}
Hence, for $A\neq 0,$ 
\begin{eqnarray}
e^\alpha \equiv F(\omega )=\frac 1{\sqrt{2}A}[1-2AB-\sqrt{1-4AB}\tanh [\mu 
\sqrt{1-4AB}(\omega -\omega _0)]]^{\frac 12},
\end{eqnarray}
where $\omega _0$ and $\mu >0$ are constants, $4AB\leq 1,$ and $\tanh $ can
be replaced by $\coth $.

For $A=0,$ $F(\omega )$ is found to be 
\begin{eqnarray}
e^\alpha \equiv F_{\pm }(\omega )=[B^2+e^{\pm 2\mu (\omega -\omega
_0)}]^{\frac 12}.
\end{eqnarray}
If$\ B=0$ in (14) then $\beta \equiv 0$ and the solution reduces to a
diagonal metric of Einstein-Rosen form [1].

The behaviour of $R(\xi ,z)$ is needed to complete the solution and is
determined by (3). We make a globally timelike choice for $R$ to obtain
cosmological solutions. Using the coordinate freedom, the general form is 
\begin{equation}
R(\xi ,z)=\xi .
\end{equation}
With this choice the constraint equations for $\psi (\xi ,z)$ admit a first
integral, and (6)-(7) become 
\begin{eqnarray}
\dot \psi &=&-\frac 1{2\xi }+\frac 12\xi [\dot \beta ^2+\beta ^{\prime
}{}^{\;\;2}+(\dot \alpha ^2+\alpha ^{\prime }{}^2)\cosh ^2\beta ], \\
\psi ^{\prime } &=&\xi [\dot \beta \beta ^{\prime }+\dot \alpha \alpha
^{\prime }\cosh ^2\beta ].
\end{eqnarray}
Substituting $F(\omega )$ from (13) or (14) into (16)-(17) yields 
\begin{eqnarray}
\dot \psi &=&-\frac 1{2\xi }+\frac 12W^2\xi [\dot \omega ^2+\omega ^{\prime
}{}^2], \\
\psi ^{\prime } &=&W^2\xi \dot \omega \omega ^{\prime },
\end{eqnarray}
where $W^2\equiv \mu ^2(1-4AB),$ while equation (11) becomes 
\begin{eqnarray}
\ddot \omega +\frac 1\xi \dot \omega -\omega ^{\prime \prime }=0.
\end{eqnarray}
This is solved by 
\begin{eqnarray}
\omega (\xi ,z)=\omega _0+\alpha _0\ln \xi +\Gamma z+\sum_{n=1}^\infty \cos
[n(z-z_n)][A_nJ_0(n\xi )+B_nN_0(n\xi )],
\end{eqnarray}
where $A_n$, $B_n$, $\omega _0$, $\alpha _0$, $z_n$, $\Gamma $ are arbitrary
constants (so we may set $\mu =1)$; $J_0$ and $N_0$ are zero-order Bessel
and Neumann functions.

Define a function, $G(\xi ,z),$ which solves the following equations: 
\begin{eqnarray}
\dot G &=&\frac 12\xi (\dot \omega ^2+\omega ^{\prime }{}^2), \\
G^{\prime } &=&\xi \dot \omega \omega ^{\prime }.
\end{eqnarray}
Eqns. (18) and (19) can be combined into a total differential

\[
d\psi =-\frac 12d\ln \xi +W^2dG 
\]
and so 
\begin{eqnarray}
\psi (\xi ,z)=-\frac 12\ln \xi +W^2G(\xi ,z)+M,
\end{eqnarray}
where $M$ is a constant.

Charach {\it et al} [7] have studied a system of equations very similar to
(16)-(20) in a situation that corresponds to $\Gamma =0$. Generalising to $%
\Gamma \neq 0,$ we find from (22)-(23),

\begin{eqnarray*}
G(\xi ,z) &=&\omega _0+\Gamma \alpha _0z+\frac 12\alpha _0^2\ln \xi +\frac
14\Gamma ^2\xi ^2 \\
&&\ \ \ \ \ \ \ \ \ \ \ \ \ \ +\alpha _0\sum_{n=1}^\infty \cos
[n(z-z_n)][A_nJ_0(n\xi )+B_nN_0(n\xi )] \\
&&\ \ \ \ \ \ \ \ \ \ \ -\Gamma \xi \sum_{n=1}^\infty \sin
[n(z-z_n)][A_nJ_1(n\xi )+B_nN_1(n\xi )] \\
&&\ \ \ \ \ \ \ +\frac 14\xi ^2\sum_{n=1}^\infty n^2([A_nJ_0(n\xi
)+B_nN_0(n\xi )]^2+[A_nJ_1(n\xi )+B_nN_1(n\xi )]^2) \\
&&\ \ \ \ \ \ \ \ -\frac 12\xi \sum_{n=1}^\infty n\cos
^2[n(z-z_n)]\{A_n^2J_0(n\xi )J_1(n\xi ) \\
&&\ \ \ \ \ +A_nB_n[N_0(n\xi )J_1(n\xi )+J_0(n\xi )N_1(n\xi )]+B_n^2N_0(n\xi
)N_1(n\xi )\} \\
&&\ \ \ \ +\frac 12\xi \sum_{n=1}^\infty \sum_{m=1,n\neq m}^\infty \frac{nm}{%
n^2-m^2}\{\sin [n(z-z_n)]\sin [m(z-z_m)][nU_{nm}^{(0)}(\xi
)-mU_{nm}^{(1)}(\xi )] \\
&&\ \ \ +\cos [n(z-z_n)]\cos [m(z-z_m)][mU_{nm}^{(0)}(\xi
)-nU_{nm}^{(1)}(\xi )]\} \\
&&
\end{eqnarray*}
where 
\begin{eqnarray}
&&  \nonumber \\
U_{nm}^{(0)}(\xi ) &\equiv &A_nA_mJ_1(n\xi )J_0(n\xi )+B_nB_mN_0(m\xi
)N_1(n\xi )+2A_nB_mJ_1(n\xi )N_0(m\xi ),
\end{eqnarray}
\begin{eqnarray}
U_{nm}^{(1)}(\xi )\equiv A_nA_mJ_0(n\xi )J_0(m\xi )+B_nB_mN_0(n\xi )N_1(m\xi
)+2A_nB_mJ_0(n\xi )N_1(m\xi ).  \nonumber
\end{eqnarray}
The full solution for the four unknown metric functions $\alpha ,\beta ,\psi
,$and $R$ is given by eqns. (10), (13) or (14), (15), (21), (24), and (25).
In order to understand the evolutionary behaviour of the solution we examine
small and large values of $n\xi $. The initial curvature singularity of the
universe lies at $\xi =0$ and its expansion continues forever as $\xi
\rightarrow \infty .$ The product $n\xi $ can be interpreted as the ratio of
the horizon distance in $z$-direction ($\triangle z=\int d\xi )$ to some
coordinate wavelength proportional to $n^{-1}$. The gravitational-wave
inhomogeneities lie outside the horizon and evolve quasi-homogeneously when $%
n\xi <1$, but are eventually encompassed by the horizon and oscillate when $%
n\xi >1,$ (density inhomogeneities behave similarly in non-vacuum models
when the sound speed equals the speed of light, [8])$.$ Here, we are
interested in the overall behaviour of the cosmological model rather than
that of particular modes of inhomogeneity and so $n$ is assumed to be of
order $1$. We examine the solution as $\xi \rightarrow 0$ and $\xi
\rightarrow \infty $. It is convenient to express $g_{ab}$ in terms of $%
F(\omega ),$ since

\begin{eqnarray}
g_{ab}=\xi (1-2AB-A^2F^2-B^2F^{-2})^{-\frac 12}\left( 
\begin{array}{lr}
F & AF+BF^{-1} \\ 
AF+BF^{-1} & F^{-1}
\end{array}
\right) .
\end{eqnarray}

{\bf The Limit }$n\xi \ll 1$

There are two cases: $A\neq 0$ and $A=0$.

{\bf (I) {\it The Case} $A\neq 0:$}

The asymptotic form of $\omega (\xi ,z)$ is

\begin{eqnarray}
\omega (\xi ,z) &\rightarrow &\omega _0+\Gamma z+\sum_{n=1}^\infty \cos
[n(z-z_n)][A_n+\frac 2\pi B_n\ln \;n]  \nonumber \\
&&\ \ \ \ +\ln \xi [\alpha _0+\frac 2\pi \sum_{n=1}^\infty B_n\cos
[n(z-z_n)]].  \nonumber
\end{eqnarray}

This implies that $\omega \rightarrow -\infty $ as $\xi \rightarrow 0$.
Hence, in this limit we have 
\[
F(\omega )\rightarrow \frac 1{\sqrt{2}A}[1-2AB+\sqrt{1-4AB}]^{\frac
12}\equiv F_0, 
\]

\begin{eqnarray*}
g_{ab}\rightarrow \xi \left( 
\begin{array}{lr}
F_0 & AF_0+BF_0^{-1} \\ 
AF_0+BF_0^{-1} & F_0^{-1}
\end{array}
\right) .
\end{eqnarray*}
Since

\begin{eqnarray}
G(\xi ,z) &\rightarrow &\omega _0+\Gamma \alpha _0z+\alpha
_0\sum_{n=1}^\infty \cos [n(z-z_n)][A_n+\frac 2\pi B_n\ln \;n]  \nonumber \\
&&\ \ +\ln \xi [\frac 12\alpha _0^2+\alpha _0\frac 2\pi \sum_{n=1}^\infty
B_n\cos [n(z-z_n)]]  \nonumber
\end{eqnarray}
we have $e^\psi \rightarrow e^{\alpha _1(z)}\xi ^{\alpha _2(z)}$ where 
\begin{equation}
\alpha _1(z)\equiv M+W^2\omega _0+W^2\Gamma \alpha _0z+\alpha
_0W^2\sum_{n=1}^\infty \cos [n(z-z_n)][A_n+\frac 2\pi B_n\ln \;n]
\end{equation}

\begin{equation}
\alpha _2(z)\equiv \frac{W^2}2\alpha _0^2-\frac 12+\alpha _0W^2\frac 2\pi
\sum_{n=1}^\infty B_n\cos [n(z-z_n)]
\end{equation}
The change to proper time $t$ is given by $dt^2=e^\psi d\xi ^2$ and so $%
t\sim \xi ^{\frac{\alpha _2(z)+2}2}$. It is convenient to choose $\alpha
_2(z)$ such that the zero points of $\xi $ and $t$ coincide; that is, with $%
\alpha _2(z)>-2$. We see that $g_{ab}$ has off-diagonal components and is
linear in $\xi $ as $\xi \rightarrow 0$. As $\xi $ $\rightarrow 0$ the
universe shrinks along the $x$- and $y$-axes, but the behaviour along the $z$%
-axis depends on the overall sign of $\alpha _2(z)$. In the range $-2<\alpha
_2(z)<0$ the expansion along the $z$-axis blows up and produces a spindle
singularity. However, choosing $\alpha _2(z)$ to be positive results in a
universe shrinking in all direction to a point-like singularity. In general,
the oscillatory character of $\alpha _2(z)$ will lead to a complicated
oscillatory spatial structure with some regions expanding along the $z$%
-direction whilst others implode according to the relative magnitudes of the
constants $\alpha _0,W,$ and $B_n$.

{\bf (II) {\it The Case }$A=0:$}

The interesting case arises from $F=$ $F_{-}$ in (14) since $F_{+}$ $%
\rightarrow B$ as $\omega \rightarrow -\infty ,$ which reduces to case (I).
However, $F_{-}$ $\rightarrow \infty $ as $\omega \rightarrow -\infty $ with 
$F_{-}(\omega )\sim \xi ^{-\alpha _3(z)}$ where

\begin{equation}
\alpha _3(z)\equiv \mu [\alpha _0+\frac 2\pi \sum_{n=1}^\infty B_n\cos
[n(z-z_n)]]
\end{equation}
and so 
\begin{eqnarray*}
g_{ab}\sim \left( 
\begin{array}{lr}
\xi ^{1-\alpha _3(z)} & B\xi ^{1+\alpha _3(z)} \\ 
B\xi ^{1+\alpha _3(z)} & \xi ^{1+\alpha _3(z)}
\end{array}
\right) .
\end{eqnarray*}

Neglecting the off-diagonal components as $\xi \rightarrow 0,$ and
re-expressing $\xi $ in terms of proper time $t,$ in regions where $\alpha
_3(z)>0$ the approach to the singularity has the form 
\[
(t^{2p_1},t^{2p_2},t^{2p_3})=(t^{\frac{2(1-\alpha _3)}{\alpha _2(z)+2}},t^{%
\frac{2(1+\alpha _3)}{\alpha _2(z)+2}},t^{\frac{2\alpha _2(z)}{\alpha _2(z)+2%
}}),
\]
so $\sum_{i=1}^3p_i=1$ and $\sum_{i=1}^3p_i^2\geq 1.$\\ 

\vspace{0.5 cm}

{\bf The Limit }$n\xi \gg 1$

{\bf (I) The Case $A\neq 0:$}

The asymptotic form of $\omega $ is

\begin{eqnarray}
\omega (\xi ,z) &\rightarrow &\omega _0+\Gamma z+\alpha _0\ln \xi  \nonumber
\\
&&\ \ \ +\xi ^{-\frac 12}\sum_{n=1}^\infty \left( \frac 2{\pi n}\right)
^{\frac 12}\cos [n(z-z_n)][A_n\cos (n\xi -\frac \pi 4)+B_n\sin (n\xi -\frac
\pi 4)]
\end{eqnarray}
so $\omega \rightarrow \infty $ as $\xi \rightarrow \infty .$ Now

\[
F(\omega )\rightarrow \frac 1{\sqrt{2}A}[1-2AB-\sqrt{1-4AB}]^{\frac
12}\equiv F_0, 
\]
\begin{eqnarray}
G(\xi ,z) &\rightarrow &\omega _0+\Gamma \alpha _0z+\frac 12\alpha _0^2\ln
\xi  \nonumber \\
&&\ \ +\xi \frac 1{2\pi }\sum_{n=1}^\infty n[A_n^2+B_n^2]+\frac 14\Gamma
^2\xi ^2,  \nonumber
\end{eqnarray}
$\ \ $so that 
\begin{eqnarray*}
g_{ab}\rightarrow \xi \left( 
\begin{array}{lr}
F_0 & AF_0+BF_0^{-1} \\ 
AF_0+BF_0^{-1} & F_0^{-1}
\end{array}
\right)
\end{eqnarray*}
and $e^\psi \rightarrow e^{\gamma _1(z)}\xi ^{\gamma _2}e^{\gamma _3\xi
+\gamma _4\xi ^2}$ as $\xi \rightarrow \infty ,$ where the four $\gamma _i$
are defined by 
\begin{equation}
\{\gamma _1(z),\gamma _2\}\equiv \{M+W^2\omega _0+W^2\Gamma \alpha _0z,\
\frac 12W^2\alpha _0^2-\frac 12\},
\end{equation}
\[
\]
\begin{equation}
\{\gamma _3,\gamma _4\}\equiv \{\frac{W^2}{2\pi }\sum_{n=1}^\infty
n[A_n^2+B_n^2],\frac{W^2}4\Gamma ^2\}
\end{equation}
\[
\ 
\]
The $\xi (t)$ relation cannot generally be integrated in this case since

\begin{eqnarray*}
t\sim \int \xi ^{\frac{\gamma _2}2}e^{\frac{\gamma _3\xi +\gamma _4\xi ^2}2\
}d\xi
\end{eqnarray*}
However, if the choice $\gamma _2=2$ is made then $t\sim \exp \{\frac
12\gamma _4(\xi +\Delta )^2\}+t_0$, $t_0$ constant,where $\triangle \equiv
\gamma _3/(2\gamma _4)\ $. If $\gamma _4>0$, $t$ is a monotonically growing
function of $\xi $ for $\xi >0$, so in this particular case the behaviour of
the solution in $\xi $ time reflects the behaviour in proper time $t$. If $%
\Gamma =\gamma _4=0$ then $t\sim \xi ^{\gamma _2/2}\ \exp \{\gamma _3\xi
/2\}.\ $In the limit $\xi \rightarrow \infty $ anisotropic expansion is
found: the expansion rates along the $x$- and $y$-axes equalise but differ
from that in the $z$-direction. There can be no genuine isotropic limit
because there is no flat vacuum Friedmann universe.

{\bf (II) The Case $A=0:$}

We have $F_{-}(\omega )\rightarrow B$ as $\omega \rightarrow \infty \ $(this
is equivalent to Case (I)), whereas $F_{+}(\omega )\rightarrow \infty .$
Neglecting the oscillatory part of $\omega (\xi ,z),$ the asymptotic
behaviour of $F_{+}$ is $F_{+}(\omega )\sim \xi ^{\mu \alpha _0}$, and hence 
\begin{eqnarray*}
g_{ab}\sim \left( 
\begin{array}{lr}
\xi ^{1+\mu \alpha _0} & B\xi ^{1-\mu \alpha _0} \\ 
B\xi ^{1-\mu \alpha _0} & \xi ^{1-\mu \alpha _0}
\end{array}
\right) .
\end{eqnarray*}
Recalling that $e^\psi \sim \xi ^{\gamma _2}\exp \{\gamma _3\xi +\gamma
_4\xi ^2\},$ and neglecting the off-diagonal components of $g_{ab},$ we see
that this describes an anisotropic universe at late times.

Our solutions are of particular interest for the problem of determining the
meaning and characterisation of a general solution of the Einstein equations
near a cosmological singularity [9]. It was argued in [6] that a metric of
the form (1)-(2) is a leading-order approximation to a 'general'
cosmological solution of Einstein's equations characterised by four
independent arbitrary functions of three space variables on a spacelike
hypersurface of constant time. This requires that derivatives with respect
to the variables $x^a$ to be smaller than those in the $z$ and $t$
variables, so that the $x^a$ appear only as functions of integration, and
that the metric component $g_{a3}$(omitted from the metric (1)-(2))
be smaller than $g_{zz}$. These requirements are met if $g_{zz}/g_{ab}\ll 1,$
[6]. When this condition holds our solution lays claim to being a
leading-order approximation to part of a general cosmological solution of
Einstein's equations [10]. It is an inhomogeneous generalisation of a single
cycle of Mixmaster oscillations during which one axis $(g_{zz})$ changes
monotonically whilst the other two oscillate [11]. Now, in the limit $n\xi
\ll 1,$ when $A\neq 0$ we have $g_{zz}/g_{ab}\sim \xi ^{\alpha _2-1}\sim
t^{2(\alpha _2-1)/(\alpha _2+2)}\ \ \ll 1$ only if $\alpha _2>1;$ however,
when $A=0$ we have $g_{zz}/g_{xx}\sim \xi ^{\alpha _2+\alpha _3-1}$ and $%
g_{zz}/g_{xy}\sim g_{zz}/g_{yy}\sim \xi ^{\alpha _2-\alpha _3-1}$, so $%
g_{zz}/g_{ab}\ll 1$ requires $\alpha _2>1$ and $\left| \alpha _3\right| $ $%
<\alpha _2-1.$ The $\alpha _{i}$ are defined by (27)-(29). When $n\xi
\gg 1$ we have, for $A\neq 0,$ that $g_{zz}/g_{ab}\ \sim \xi ^{\gamma
_2-1}\exp \{\gamma _3\xi +\gamma _4\xi ^2\}$. The $\gamma _i$ are defined by
(31)-(32). Since $\gamma _{3}$ and $\gamma _4$ are positive we have $%
g_{zz}/g_{ab}\gg 1$ in this case, and our solution is not a leading-order
approximation to the general solution when $\xi \rightarrow \infty $. When $%
A=0$ for $n\xi \gg 1,$ we have $g_{zz}/g_{xx}\sim \xi ^{\gamma _2-1-\mu
\alpha _0}\exp \{\gamma _3\xi +\gamma _4\xi ^2\}$ and $g_{zz}/g_{xy}\sim \xi
^{\gamma _2-1+\mu \alpha _0}\exp \{\gamma _3\xi +\gamma _4\xi ^2\}$. Again,
as in the $A\neq 0$ case, the solution cannot be the leading-order
approximation to a general (4-function) solution. Thus for certain parameter
choices and over certain intervals of time our solutions provide new
information about a general solution of Einstein's equations near a
cosmological singularity. \\ 

\vspace{0.5 cm}

{\bf Acknowledgements} JDB is supported by the PPARC and KEK by the German
National Scholarship Foundation.\\

\vspace{0.5 cm}

{\bf References}

[1] A. Einstein and N. Rosen, J. Franklin Inst.{\bf \ 223}, 43 (1937); A.S.
Kompanyeets, Sov. Phys. JETP {\bf 7}, 659 (1958).

[2] M. Carmeli, Ch. Charach, and S. Malin, Phys. Rep. {\bf 76,} 79 (1981);
A. Krasi\'nski, {\it Physics in an Inhomogeneous Universe,} (N. Copernicus
Astronomical Center, Warsaw, 1993).

[3] P. Szekeres, J. Math. Phys. {\bf 13,} 286 (1972); J.B. Griffiths, {\it %
Colliding Plane Gravitational Waves in General Relativity, }(Oxford UP,
Oxford, 1991).

[4] K. Tomita, Prog. Theor. Phys. {\bf 59,} 1150 (1978).

[5] R. Gowdy, Phys. Rev. Lett. {\bf 27}, 827 (1971); Ann. Phys. (NY) {\bf 83}%
, 203 (1974).

[6] V. A. Belinskii and I. M. Khalatnikov, Sov. Phys. JETP {\bf 30,} 1174
(1970); Sov. Phys. JETP {\bf 32,} 169 (1971).

[7] Ch. Charach, Phys. Rev. D{\bf 19,} 3516 (1979); M. Carmeli and Ch.
Charach, Phys. Lett. A {\bf 75,} 333 (1980); Ch. Charach and S. Malin, Phys.
Rev. D {\bf 21,} 3284 (1980).

[8] J.D. Barrow and K.E. Kunze, paper in preparation.

[9] J.D. Barrow and F.J. Tipler, Phys. Rep. {\bf 56}, 371 (1979).

[10] Our exact solution is generally determined by $A,B,\omega (\xi ,z)$ and 
$G(\xi ,z)$. The class of functions represented by the series expansions in
(21) and (25) is restricted.

[11] Chaotic behaviour is not displayed by these oscillations within a
cycle, only by cycle-to-cycle transformations of the non-oscillatory axis,
see J.D. Barrow, Phys. Rep.{\bf \ 85}, 1 (1982).

\end{document}